\documentclass[aps,prl,twocolumn,superscriptaddress,floatfix,amsmath,longbibliography,10pt]{revtex4-2} 

\usepackage[normalem]{ulem}
\usepackage{bm} 
\usepackage{amssymb}
\usepackage{amsfonts} 
\usepackage{amsmath} 

\usepackage{graphicx} 

\usepackage{xcolor} 
 
\usepackage{siunitx}
\DeclareSIUnit{\dBm}{dBm}
\DeclareSIUnit{\Oe}{Oe}

\usepackage[pdftex,colorlinks=true,allcolors=blue,breaklinks=true]{hyperref}  

\usepackage[utf8]{inputenc} 
\usepackage{textgreek}
 
\definecolor{g-blue}{rgb}{0.83,0.95,1}
\definecolor{g-yellow}{rgb}{1,1,0.7}
\definecolor{g-green}{rgb}{0.9,1,0.9}
\definecolor{green}{rgb}{0,0.6,0}
\definecolor{cyan}{rgb}{0,0.7,0.7}
\definecolor{black}{rgb}{0,0,0}
\definecolor{grey}{rgb}{0.4,0.4,0.4}
\definecolor{nature-blue}{rgb}{0.0,0.200,0.500}

\def \ed {\end{document}}
\def\Fbox#1{\vskip1ex\hbox to 8.5cm{\hfil\fboxsep0.3cm\fbox{%
		\parbox{8.0cm}{#1}}\hfil}\vskip1ex\noindent}  

\let \= \equiv  
\let\*\cdot 
\let\^\widehat 
\let\-\overline

\def\1{\bm1}

\def\<{\left\langle}    \def\>{\right\rangle}
\def\[ {\left[}         \def\]{\right]}

\begin{document}

\title{Bulk-mediated reflection of chirality-protected surface spin waves}


\author{Vitaliy I. Vasyuchka}
 \thanks{These authors contributed equally to this work}
 \affiliation{Fachbereich Physik and Landesforschungszentrum OPTIMAS, Rheinland-Pf\"alzische Technische Universit\"at Kaiserslautern-Landau, 67663 Kaiserslautern, Germany}

\author{Florin Ciubotaru}
 \thanks{These authors contributed equally to this work}
 \affiliation{imec, B-3001, Leuven, Belgium}

\author{Andrii V. Chumak}
 \affiliation{Faculty of Physics, University of Vienna, 1090 Vienna, Austria} 

\author{Burkard Hillebrands}
 \affiliation{Fachbereich Physik and Landesforschungszentrum OPTIMAS, Rheinland-Pf\"alzische Technische Universit\"at Kaiserslautern-Landau, 67663 Kaiserslautern, Germany}

\author{Alexander A. Serga}
 \email{serha@rptu.de}
 \affiliation{Fachbereich Physik and Landesforschungszentrum OPTIMAS, Rheinland-Pf\"alzische Technische Universit\"at Kaiserslautern-Landau, 67663 Kaiserslautern, Germany}

\date{\today}

\begin{abstract}  
Surface spin waves of the Damon--Eshbach type exhibit intrinsically nonreciprocal transport properties with a chiral dynamical field structure that localizes counterpropagating waves at opposite film surfaces. Such chirality has been predicted to suppress direct backscattering in thin films within frequency ranges free of bulk modes. However, how chirality influences reflection in thicker three-dimensional magnetic media, where a dense spectrum of bulk excitations overlaps with surface waves, remains unclear. Here we demonstrate that, in micrometer-thick yttrium iron garnet films, reflection of the chiral Damon--Eshbach wave from the boundary of the magnetic medium is accompanied by excitation of spatially localized thickness-quantized bulk modes, whereas reciprocal backward-volume waves reflect nearly elastically. Brillouin light scattering spectroscopy, infrared thermography, and micromagnetic simulations reveal standing bulk excitations at the reflecting boundary and quantify the associated magnon energy accumulation and dissipation. These results identify bulk-mode excitations as the physical pathway enabling reversal of chirally localized surface waves in thick films, thereby defining the limits of chirality-based backscattering immunity and providing a general framework for wave transport in nonreciprocal magnetic media.

\end{abstract}

\maketitle

Chirality has emerged as a unifying concept across modern magnetism and spintronics, shaping both the structure of magnetic textures and the dynamics of spin excitations~\cite{Yu2023}.
In magnetic systems, the intrinsic breaking of time-reversal symmetry encoded in the Landau--Lifshitz dynamics provides a natural basis for chiral effects.
Magnetodipolar interactions can lift the symmetry between counterpropagating spin-wave states, giving rise to nonreciprocal modes with asymmetric dynamical magnetization and field profiles and enabling the formation of chiral spin waves via hybridization of modes with opposite parity~\cite{Trevillian2024}. 
In two-dimensional magnonic crystals, the dipolar coupling can produce topologically protected chiral edge states residing within bulk band gaps~\cite{Shindou2013}. 
In continuous magnetic films, a prominent realization of magnetodipolar chirality is the Damon--Eshbach magnetostatic surface wave~\cite{Damon1961, Pirro2021}, which exhibits intrinsically nonreciprocal propagation with asymmetric surface localization and a handed dynamical field structure across the film thickness. 
Recent studies have suggested that the properties of this surface mode can be understood in terms of its chiral and topological character~\cite{Yamamoto2019}.\looseness=-1

\begin{figure}[t]
\centering  
\includegraphics[width=1.05\columnwidth]{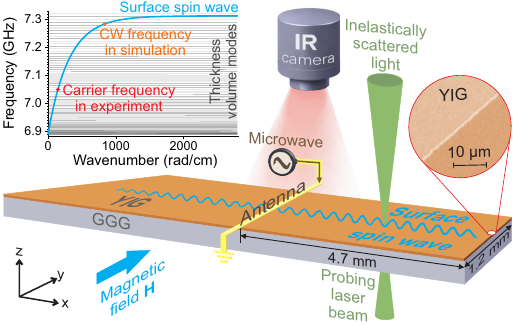}
\caption{Experimental geometry, spin-wave spectrum, and measurement scheme.
Spin waves are excited in an in-plane magnetized, single-crystalline \qty{16}{\micro\meter}-thick YIG waveguide by a microwave stripline antenna and propagate toward a reflecting edge oriented perpendicular to the waveguide axis.
The right inset shows a micrograph of the cleaved reflecting edge forming the spin-wave mirror.
The shown external magnetic field corresponds to the Damon--Eshbach geometry; the reference reciprocal backward-volume configuration is realized by aligning the external magnetic field along the waveguide.
Spin-wave intensity is measured by space- and time-resolved Brillouin light scattering (BLS) spectroscopy, while dissipated magnon energy is independently mapped by infrared thermography.
\looseness=-1
        }
        \label{f:setup}
\end{figure}

Indeed, micromagnetic simulations have demonstrated suppression of direct backscattering of Damon--Eshbach waves from surface topographic defects in thin films within frequency ranges where bulk modes are absent for collinear propagation~\cite{Mohseni2019}. 
However, these frequency windows do not constitute a true two-dimensional bulk band gap, since bulk excitations remain available if the direction of the scattered wave vector is changed---unlike the fully gapped topological edge states in two-dimensional magnonic crystals~\cite{Shindou2013}. Moreover, the considered geometry involves scattering from a shallow surface defect rather than from the termination of the magnetic medium. 
In thicker magnetic films, where a dense continuum of thickness-quantized bulk modes overlaps spectrally with surface waves (see the spin-wave spectrum in Fig.\,\ref{f:setup}), spectral isolation of the surface mode is absent, and it remains unclear to what extent chirality can influence reflection under such fully gapless spectral conditions.\looseness=-1

Here, we demonstrate that reflection of the chiral Damon--Eshbach wave from the edge of a magnetic waveguide involves excitation of spatially localized thickness-quantized volume modes, which provide a pathway for relocating the surface wave across the film thickness without direct backscattering.
Using Brillouin light scattering (BLS) spectroscopy, micromagnetic simulations, and infrared thermography, we directly identify this bulk-mediated reflection pathway and quantify the accompanying magnon energy accumulation and dissipation.\looseness=-1

Figure\,\ref{f:setup} schematically illustrates the sample geometry and the experimental techniques employed in this work. In our experiments, we investigated the reflection of magnetostatic spin waves from the end of a spin-wave waveguide prepared from a single-crystalline yttrium iron garnet film (YIG, $\mathrm{Y_3Fe_5O_{12}}$) of \qty{16}{\micro\meter} thickness grown on a gadolinium gallium garnet (GGG) substrate. 
The waveguide is \qty{1.2}{\milli\meter} wide and \qty{18}{\milli\meter} long. 
One end of the YIG stripe is cleaved perpendicular to the waveguide axis to form a well-defined reflecting boundary with minimal edge damage and reduced defect-induced scattering. The opposite end is tapered at \qty{45}{\degree} to suppress spurious backscattering processes.
The spin waves are excited by a microwave-driven stripline antenna of \qty{50}{\micro\meter} width placed across the waveguide \qty{4.7}{\milli\meter} apart from its right-angled end. The antenna is mounted on a narrow dielectric bridge, providing optical access to the sample from both sides for BLS measurements. The YIG stripe was magnetized in-plane by a bias field $H = \qty{1750}{\Oe}$ either across or along its long axis. The first case corresponds to the excitation and propagation of a nonreciprocal magnetostatic surface spin wave (MSSW, Damon--Eshbach wave) and the second one to a reference case of the reciprocal backward volume magnetostatic spin wave (BVMSW) \cite{Schneider2008, Kostylev2013, Serha2022, Pirro2021}. 
In both cases, the spatio-temporal spin-wave dynamics was measured by means of space- and time-resolved BLS spectroscopy \cite{Buettner2000} in forward-scattering geometry, where the light inelastically scattered by spin waves is collected after transmission through the film and directed to a multi-pass Fabry--P\'erot interferometer for frequency and intensity analysis \cite{Dunagin2025}.
\looseness=-1

\begin{figure}[b]
\centering  
\includegraphics[width=1.00\columnwidth]{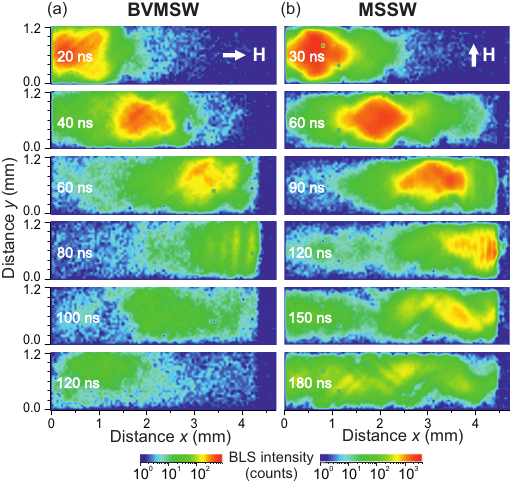}
\caption{Time-resolved spatial intensity maps of incident and reflected spin-wave packets in the YIG waveguide. 
(a) Reciprocal backward-volume waves (BVMSW) retain their shape upon reflection, consistent with elastic scattering from the edge.
(b) Chiral Damon--Eshbach surface waves (MSSW) exhibit strong distortion and elongation of the reflected packet, consistent with bulk-mediated reflection.
Maps show the BLS intensity of \qty{20}{\nano\second} pulses at successive delay times in false color logarithmic scale; $x=0$ marks the antenna position.}
        \label{f:BLS}
\end{figure}

The time-resolved BLS maps of the propagating wave packets are shown in Fig.~\ref{f:BLS}. To separate incident and reflected signals, the excitation was performed using \qty{20}{\nano\second}-long microwave pulses at \qty{6.89}{\giga\hertz} (BVMSW) and \qty{7.05}{\giga\hertz} (MSSW), with a repetition period of \qty{800}{\nano\second}. 
In both cases, the carrier frequency was approximately \qty{150}{\mega\hertz} away from the uniform precession frequency (see, e.g., Fig.\,\ref{f:setup}), ensuring a match between the spin-wave spectrum and the Fourier spectrum of the short input pulses.
With this choice of operating frequencies, the efficiency of spin wave excitation was found to be relatively low, and the transition to nonlinear regimes of spin-wave packet propagation, which could manifest itself in the formation of solitons, bullets, their nonlinear decay, or even collapse \cite{Sulymenko2018}, was not observed even at fairly high peak input powers.
Because of the different excitation efficiencies of the two modes, the applied powers were adjusted to \qty{630}{\milli\watt} for BVMSW and \qty{100}{\milli\watt} for MSSW.

Both wave types propagate from the antenna to the waveguide end without qualitative differences. 
Their reflection behavior, however, differs markedly. 
The BVMSW packet retains its shape upon reflection, and the oscillatory intensity near the edge matches the carrier wavelength, consistent with interference of the incident and reflected waves and elastic scattering at the boundary [Fig.~\ref{f:BLS}(a)].

For MSSWs, a clear interference pattern also develops near the reflecting edge; nevertheless, the reflected packet becomes strongly distorted and elongated [Fig.~\ref{f:BLS}(b)]. 
In addition, transient intensity minima and localized edge-related features appear during reflection, indicating that the scattering region develops a complex internal structure. 
Together, these observations show that, in contrast to the reciprocal case, the reflected signal does not preserve the spatio-temporal structure of the incident packet; therefore, the reflection process cannot be described as purely elastic, indicating mode conversion during reflection.\looseness=-1

To clarify the microscopic origin of this non-elastic reflection behavior, we performed micromagnetic simulations of the experimental geometry~\cite{Donahue1999}. Because modeling the full \qty{20}{\milli\meter}-long sample is computationally prohibitive, a reduced-size \qty{16}{\micro\meter}-thick YIG waveguide of \qtyproduct{4 x 1}{\milli\meter} was simulated. The structure was discretized into $400 \times 100 \times 8$ cells. Spin waves were excited by a microwave field generated by a \qty{50}{\micro\meter}-wide antenna placed \qty{2.5}{\milli\meter} from the reflecting edge, while damping boundary conditions suppressed spurious reflections from the opposite end. \looseness=-1

For the BVMSW geometry, the simulations reproduce the BLS observations: the incident and reflected packets retain nearly identical shapes, and the oscillations at the edge match the carrier wavelength, consistent with elastic reflection. 

In the MSSW geometry under the experimental conditions, substantial interference between the incident and reflected surface waves obscures possible bulk contributions. Because the MSSW amplitude decays across the film thickness approximately as $\exp(-2\pi z/\lambda)$, a reduction of the wavelength increases the surface localization and thereby decreases the spatial overlap of counterpropagating waves residing at opposite film surfaces. To exploit this effect, the excitation frequency was increased by \qty{280}{\mega\hertz} to \qty{7.285}{\giga\hertz}, shortening the wavelength from approximately \qty{435}{\micro\meter}, calculated for the experimental conditions, to \qty{75}{\micro\meter} and thus suppressing trivial interference.
Continuous excitation was further used to avoid spectral broadening of pulsed signals. These modified conditions were introduced solely to enhance the visibility of bulk modes in the simulations; experimentally, the signal at this frequency is too weak for reliable detection.

\begin{figure}[t]
\centering  
\includegraphics[width=1.00\columnwidth]{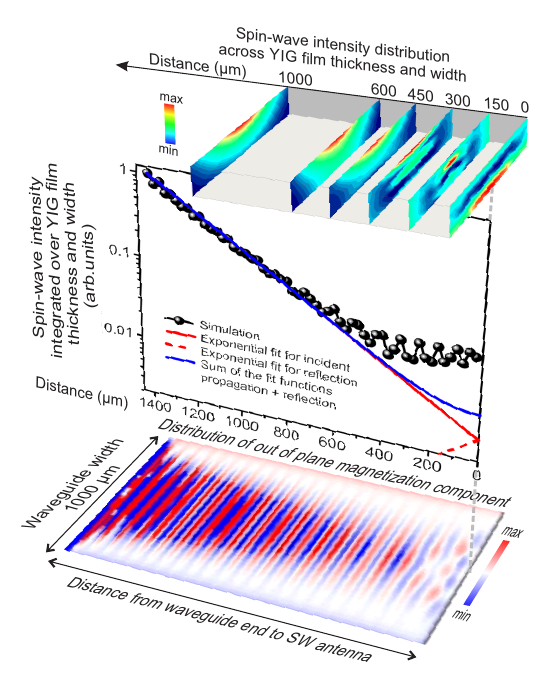}
\caption{Micromagnetic simulations revealing bulk-mediated reflection of the MSSW. 
The upper panels show cross sections of the spin-wave intensity across the film thickness and width at different distances from the reflecting edge, demonstrating the emergence of standing, thickness- and width-quantized bulk modes near the boundary. 
The middle panel plots the spin-wave energy integrated over the waveguide cross-section, showing excess energy accumulation at the edge due to these localized modes. 
The bottom panel displays the spin-wave intensity map illustrating the corresponding standing-wave pattern over the upper film surface. 
Continuous excitation at \qty{7.285}{\giga\hertz} is used to suppress interference between incident and reflected waves and resolve the bulk excitations.
        }
        \label{f:simulation_overview}
\end{figure}

Under these conditions, the simulated profiles shown in Fig.~\ref{f:simulation_overview} reveal clear standing oscillations across both the film thickness and width near the reflecting edge. Far from the boundary, the MSSW exhibits the expected exponential surface localization, whereas close to the edge, the profiles develop multiple nodes characteristic of thickness- and width-quantized bulk modes. The emergence of these localized standing modes identifies bulk excitations that mediate the reflection of the surface wave.

\begin{figure}[t]
\centering  
\includegraphics[width=1.00\columnwidth]{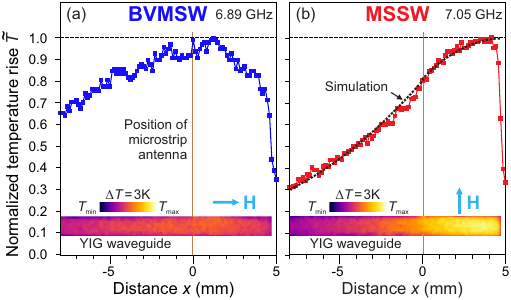}
\caption{Infrared thermography of magnon-induced energy dissipation for reciprocal BVMSWs and chiral MSSWs.
The insets show infrared images of the YIG waveguide with identical temperature scales ($\Delta T=\qty{3}{\kelvin}$).
(a) BVMSWs exhibit an almost symmetric spatial heating profile around the antenna with gradual decay in both propagation directions. 
(b) MSSWs display a strongly asymmetric heating profile, in which the temperature maximum forms near the reflecting edge instead of at the antenna, indicating localized energy accumulation associated with bulk-mediated reflection.
The plotted profiles show the normalized temperature rise $\widetilde{T}(x)$.
Vertical lines indicate the position of the microstrip antenna.
The black curve in (b) shows the COMSOL-calculated spatial temperature profile based on the bulk-mediated reflection model.
        }
        \label{f:heating}
\end{figure}

Conventional BLS is insensitive to bulk modes with wave-vector components across the film thickness \cite{Dunagin2025}. To obtain an independent signature of their excitation, we employed infrared thermography, which directly probes the spatial distribution of magnon energy dissipation. The film surface was imaged with spatial resolution of \qty{135}{\micro\meter} using an infrared camera (FLIR SC655) with a thermal sensitivity of \qty{0.05}{\kelvin}. 
The shown temperature profiles were obtained by averaging six line scans along the waveguide axis, spaced equidistantly across its width. The normalized temperature rise is defined as
$\widetilde{T}(x)=[T(x)-T_\mathrm{ref}]/[T_\mathrm{max}-T_\mathrm{ref}]$,
where $T_\mathrm{ref}$ is the spatially averaged temperature measured on the substrate outside the YIG waveguide.
To compare the spatial heating profiles independently of the absolute heating amplitude, the temperature distributions were normalized to their respective maximum values.

The measured temperature maps are shown in Fig.~\ref{f:heating}. 
For the reciprocal BVMSW case, the heating profile is nearly symmetric about the antenna position and gradually decays toward both ends of the waveguide, consistent with propagation losses of a bidirectional wave. 

A qualitatively different behavior is found for MSSWs. 
Instead of a monotonic decay, the temperature increases toward the reflecting edge and exhibits a pronounced maximum at the sample boundary [Fig.~\ref{f:heating}(b,d)], while decreasing in the opposite direction. 
Notably, the dissipation maximum is spatially separated from the excitation region, demonstrating nonlocal energy conversion during reflection.

Although nonreciprocal MSSW excitation is known to enable a unidirectional heat conveyer functionality~\cite{An2013}, that effect leads to heating localized near the antenna, where the waves are launched. 
In contrast, the temperature maximum observed here appears at the reflecting boundary. 
The weak heating reported at the tapered tip in Ref.~\cite{An2013} originates from geometric concentration of spin waves and is not applicable to the present uniform geometry. 
Therefore, the observed edge heating cannot be explained solely by the heat-conveyer mechanism.

Instead, the temperature increases precisely in the region where the simulations predict excitation of thickness- and width-quantized bulk modes.

By integrating the simulated spin-wave intensity over the film cross section and time (middle panel of Fig.~\ref{f:simulation_overview}), we obtain the spatial distribution of dissipated magnon energy along the waveguide. The calculated energy deposition decreases during propagation but rises visibly near the reflecting edge. Using this profile as a spatially distributed heating source and accounting for phonon-mediated heat transport, we performed COMSOL simulations of the resulting temperature distribution along the waveguide [black line in Fig.~\ref{f:heating}(b)]. The calculated curve is shown on an arbitrary vertical scale, since the thermal model aims to reproduce the spatial temperature profile rather than the absolute heating amplitude. 
The calculated and measured profiles exhibit the same spatial trend. 
This correspondence provides independent support for the interpretation that bulk-mode excitation and subsequent relaxation dominate the surface-wave reflection process.

Taken together, experiments and micromagnetic simulations show that reflection of magnetostatic surface waves is governed by excitation of thickness-quantized bulk modes near the boundary.
Because electrodynamic boundary conditions enforce a chiral surface-localized mode profile, elastic reversal of an MSSW would require the mode to be effectively relocated across the film thickness, thereby constituting a strong suppression mechanism for direct elastic reflection.
Excitation of bulk modes provides the additional spatial degrees of freedom required to satisfy the boundary conditions at the waveguide termination.
As a result, reflection proceeds through mode conversion involving thickness-quantized bulk modes, which redistribute the dynamic magnetization across the film thickness and enhance local dissipation at the edge.

In summary, while the edge of a magnetic medium acts as an almost perfect mirror for reciprocal backward-volume waves, reflection of chiral surface waves occurs through bulk-mode mediation rather than direct backscattering. This mechanism explains the long-standing observation that surface waves are weakly reflected by thickness modulations \cite{Chumak2009} and clarifies the limits of chirality-based backscattering protection in realistic geometries. We show that chirality-protected immunity remains robust against direct backscattering even in micrometer-thick, purely dipolar films, where reversal becomes possible only through excitation of thickness- and width-quantized bulk modes. 
These findings are directly relevant for studies of nonreciprocal spin-wave dynamics \cite{Wagle2026}, magnon-phonon conversion \cite{Holanda2018}, and energy localization in magnetic waveguides \cite{An2013}. 
They are also important for the design of spin-wave resonators \cite{Costa2021}, magnonic crystals \cite{Whitney2025}, and interference-based microwave devices, including emerging spin-wave RF platforms envisioned for 5G and 6G communication technologies \cite{Levchenko2026}.

V.I.V. and F.C. contributed equally to this work. V.I.V. and F.C. performed the measurements and, together with A.V.C. and A.A.S., analysed the data; F.C. carried out the micromagnetic simulation; V.I.V., F.C., A.V.C., and A.A.S. contributed to the experimental setup; A.A.S. wrote the manuscript; all authors discussed the results and commented on the manuscript; A.A.S., V.I.V., and B.H. planned and supervised the project.

This research was funded in part by the Deutsche Forschungsgemeinschaft (DFG, German Research Foundation) in the framework of TRR 173 -- Grant No. 268565370 ``Spin+X'' (Projects B01 and B13) and by the Austrian Science Fund (FWF) through the project MagNeuro [10.55776/PIN1434524]. The authors are grateful to V. S. Tyberkevych for fruitful discussions.


\begin{thebibliography}{21}%
	\makeatletter
	\providecommand \@ifxundefined [1]{%
		\@ifx{#1\undefined}
	}%
	\providecommand \@ifnum [1]{%
		\ifnum #1\expandafter \@firstoftwo
		\else \expandafter \@secondoftwo
		\fi
	}%
	\providecommand \@ifx [1]{%
		\ifx #1\expandafter \@firstoftwo
		\else \expandafter \@secondoftwo
		\fi
	}%
	\providecommand \natexlab [1]{#1}%
	\providecommand \enquote  [1]{``#1''}%
	\providecommand \bibnamefont  [1]{#1}%
	\providecommand \bibfnamefont [1]{#1}%
	\providecommand \citenamefont [1]{#1}%
	\providecommand \href@noop [0]{\@secondoftwo}%
	\providecommand \href [0]{\begingroup \@sanitize@url \@href}%
	\providecommand \@href[1]{\@@startlink{#1}\@@href}%
	\providecommand \@@href[1]{\endgroup#1\@@endlink}%
	\providecommand \@sanitize@url [0]{\catcode `\\12\catcode `\$12\catcode `\&12\catcode `\#12\catcode `\^12\catcode `\_12\catcode `\%12\relax}%
	\providecommand \@@startlink[1]{}%
	\providecommand \@@endlink[0]{}%
	\providecommand \url  [0]{\begingroup\@sanitize@url \@url }%
	\providecommand \@url [1]{\endgroup\@href {#1}{\urlprefix }}%
	\providecommand \urlprefix  [0]{URL }%
	\providecommand \Eprint [0]{\href }%
	\providecommand \doibase [0]{https://doi.org/}%
	\providecommand \selectlanguage [0]{\@gobble}%
	\providecommand \bibinfo  [0]{\@secondoftwo}%
	\providecommand \bibfield  [0]{\@secondoftwo}%
	\providecommand \translation [1]{[#1]}%
	\providecommand \BibitemOpen [0]{}%
	\providecommand \bibitemStop [0]{}%
	\providecommand \bibitemNoStop [0]{.\EOS\space}%
	\providecommand \EOS [0]{\spacefactor3000\relax}%
	\providecommand \BibitemShut  [1]{\csname bibitem#1\endcsname}%
	\let\auto@bib@innerbib\@empty
	\bibitem [{\citenamefont {Yu}\ \emph {et~al.}(2023)\citenamefont {Yu}, \citenamefont {Luo},\ and\ \citenamefont {Bauer}}]{Yu2023}%
	\BibitemOpen
	\bibfield  {author} {\bibinfo {author} {\bibfnamefont {T.}~\bibnamefont {Yu}}, \bibinfo {author} {\bibfnamefont {Z.}~\bibnamefont {Luo}},\ and\ \bibinfo {author} {\bibfnamefont {G.~E.}\ \bibnamefont {Bauer}},\ }\bibfield  {title} {\bibinfo {title} {Chirality as generalized spin-orbit interaction in spintronics},\ }\href {https://doi.org/10.1016/j.physrep.2023.01.002} {\bibfield  {journal} {\bibinfo  {journal} {Phys. Rep.}\ }\textbf {\bibinfo {volume} {1009}},\ \bibinfo {pages} {1–115} (\bibinfo {year} {2023})}\BibitemShut {NoStop}%
	\bibitem [{\citenamefont {Trevillian}\ and\ \citenamefont {Tyberkevych}(2024)}]{Trevillian2024}%
	\BibitemOpen
	\bibfield  {author} {\bibinfo {author} {\bibfnamefont {C.}~\bibnamefont {Trevillian}}\ and\ \bibinfo {author} {\bibfnamefont {V.}~\bibnamefont {Tyberkevych}},\ }\bibfield  {title} {\bibinfo {title} {Formation of chirality in propagating spin waves},\ }\href {https://doi.org/10.1038/s44306-024-00026-3} {\bibfield  {journal} {\bibinfo  {journal} {npj Spintronics}\ }\textbf {\bibinfo {volume} {2}},\ \bibinfo {pages} {23} (\bibinfo {year} {2024})}\BibitemShut {NoStop}%
	\bibitem [{\citenamefont {Shindou}\ \emph {et~al.}(2013)\citenamefont {Shindou}, \citenamefont {Matsumoto}, \citenamefont {Murakami},\ and\ \citenamefont {Ohe}}]{Shindou2013}%
	\BibitemOpen
	\bibfield  {author} {\bibinfo {author} {\bibfnamefont {R.}~\bibnamefont {Shindou}}, \bibinfo {author} {\bibfnamefont {R.}~\bibnamefont {Matsumoto}}, \bibinfo {author} {\bibfnamefont {S.}~\bibnamefont {Murakami}},\ and\ \bibinfo {author} {\bibfnamefont {J.-i.}\ \bibnamefont {Ohe}},\ }\bibfield  {title} {\bibinfo {title} {Topological chiral magnonic edge mode in a magnonic crystal},\ }\href {https://doi.org/10.1103/physrevb.87.174427} {\bibfield  {journal} {\bibinfo  {journal} {Phy. Rev. B}\ }\textbf {\bibinfo {volume} {87}},\ \bibinfo {pages} {174427} (\bibinfo {year} {2013})}\BibitemShut {NoStop}%
	\bibitem [{\citenamefont {Damon}\ and\ \citenamefont {Eshbach}(1961)}]{Damon1961}%
	\BibitemOpen
	\bibfield  {author} {\bibinfo {author} {\bibfnamefont {R.}~\bibnamefont {Damon}}\ and\ \bibinfo {author} {\bibfnamefont {J.}~\bibnamefont {Eshbach}},\ }\bibfield  {title} {\bibinfo {title} {Magnetostatic modes of a ferromagnet slab},\ }\href {https://doi.org/10.1016/0022-3697(61)90041-5} {\bibfield  {journal} {\bibinfo  {journal} {J. Phys. Chem. Solids}\ }\textbf {\bibinfo {volume} {19}},\ \bibinfo {pages} {308–320} (\bibinfo {year} {1961})}\BibitemShut {NoStop}%
	\bibitem [{\citenamefont {Pirro}\ \emph {et~al.}(2021)\citenamefont {Pirro}, \citenamefont {Vasyuchka}, \citenamefont {Serga},\ and\ \citenamefont {Hillebrands}}]{Pirro2021}%
	\BibitemOpen
	\bibfield  {author} {\bibinfo {author} {\bibfnamefont {P.}~\bibnamefont {Pirro}}, \bibinfo {author} {\bibfnamefont {V.~I.}\ \bibnamefont {Vasyuchka}}, \bibinfo {author} {\bibfnamefont {A.~A.}\ \bibnamefont {Serga}},\ and\ \bibinfo {author} {\bibfnamefont {B.}~\bibnamefont {Hillebrands}},\ }\bibfield  {title} {\bibinfo {title} {Advances in coherent magnonics},\ }\href {https://doi.org/10.1038/s41578-021-00332-w} {\bibfield  {journal} {\bibinfo  {journal} {Nat. Rev. Mater.}\ }\textbf {\bibinfo {volume} {6}},\ \bibinfo {pages} {1114–1135} (\bibinfo {year} {2021})}\BibitemShut {NoStop}%
	\bibitem [{\citenamefont {Yamamoto}\ \emph {et~al.}(2019)\citenamefont {Yamamoto}, \citenamefont {Thiang}, \citenamefont {Pirro}, \citenamefont {Kim}, \citenamefont {Everschor-Sitte},\ and\ \citenamefont {Saitoh}}]{Yamamoto2019}%
	\BibitemOpen
	\bibfield  {author} {\bibinfo {author} {\bibfnamefont {K.}~\bibnamefont {Yamamoto}}, \bibinfo {author} {\bibfnamefont {G.~C.}\ \bibnamefont {Thiang}}, \bibinfo {author} {\bibfnamefont {P.}~\bibnamefont {Pirro}}, \bibinfo {author} {\bibfnamefont {K.-W.}\ \bibnamefont {Kim}}, \bibinfo {author} {\bibfnamefont {K.}~\bibnamefont {Everschor-Sitte}},\ and\ \bibinfo {author} {\bibfnamefont {E.}~\bibnamefont {Saitoh}},\ }\bibfield  {title} {\bibinfo {title} {Topological characterization of classical waves: {T}he topological origin of magnetostatic surface spin waves},\ }\href {https://doi.org/10.1103/physrevlett.122.217201} {\bibfield  {journal} {\bibinfo  {journal} {Phys. Rev. Lett.}\ }\textbf {\bibinfo {volume} {122}},\ \bibinfo {pages} {217201} (\bibinfo {year} {2019})}\BibitemShut {NoStop}%
	\bibitem [{\citenamefont {Mohseni}\ \emph {et~al.}(2019)\citenamefont {Mohseni}, \citenamefont {Verba}, \citenamefont {Br\"{a}cher}, \citenamefont {Wang}, \citenamefont {Bozhko}, \citenamefont {Hillebrands},\ and\ \citenamefont {Pirro}}]{Mohseni2019}%
	\BibitemOpen
	\bibfield  {author} {\bibinfo {author} {\bibfnamefont {M.}~\bibnamefont {Mohseni}}, \bibinfo {author} {\bibfnamefont {R.}~\bibnamefont {Verba}}, \bibinfo {author} {\bibfnamefont {T.}~\bibnamefont {Br\"{a}cher}}, \bibinfo {author} {\bibfnamefont {Q.}~\bibnamefont {Wang}}, \bibinfo {author} {\bibfnamefont {D.~A.}\ \bibnamefont {Bozhko}}, \bibinfo {author} {\bibfnamefont {B.}~\bibnamefont {Hillebrands}},\ and\ \bibinfo {author} {\bibfnamefont {P.}~\bibnamefont {Pirro}},\ }\bibfield  {title} {\bibinfo {title} {Backscattering immunity of dipole-exchange magnetostatic surface spin waves},\ }\href {https://doi.org/10.1103/physrevlett.122.197201} {\bibfield  {journal} {\bibinfo  {journal} {Phys. Rev. Lett.}\ }\textbf {\bibinfo {volume} {122}},\ \bibinfo {pages} {197201} (\bibinfo {year} {2019})}\BibitemShut {NoStop}%
	\bibitem [{\citenamefont {Schneider}\ \emph {et~al.}(2008)\citenamefont {Schneider}, \citenamefont {Serga}, \citenamefont {Neumann}, \citenamefont {Hillebrands},\ and\ \citenamefont {Kostylev}}]{Schneider2008}%
	\BibitemOpen
	\bibfield  {author} {\bibinfo {author} {\bibfnamefont {T.}~\bibnamefont {Schneider}}, \bibinfo {author} {\bibfnamefont {A.~A.}\ \bibnamefont {Serga}}, \bibinfo {author} {\bibfnamefont {T.}~\bibnamefont {Neumann}}, \bibinfo {author} {\bibfnamefont {B.}~\bibnamefont {Hillebrands}},\ and\ \bibinfo {author} {\bibfnamefont {M.~P.}\ \bibnamefont {Kostylev}},\ }\bibfield  {title} {\bibinfo {title} {Phase reciprocity of spin-wave excitation by a microstrip antenna},\ }\href {https://doi.org/10.1103/physrevb.77.214411} {\bibfield  {journal} {\bibinfo  {journal} {Phys. Rev. B}\ }\textbf {\bibinfo {volume} {77}},\ \bibinfo {pages} {214411} (\bibinfo {year} {2008})}\BibitemShut {NoStop}%
	\bibitem [{\citenamefont {Kostylev}(2013)}]{Kostylev2013}%
	\BibitemOpen
	\bibfield  {author} {\bibinfo {author} {\bibfnamefont {M.}~\bibnamefont {Kostylev}},\ }\bibfield  {title} {\bibinfo {title} {Non-reciprocity of dipole-exchange spin waves in thin ferromagnetic films},\ }\href {https://doi.org/10.1063/1.4789962} {\bibfield  {journal} {\bibinfo  {journal} {J. Appl. Phys.}\ }\textbf {\bibinfo {volume} {113}},\ \bibinfo {pages} {053907} (\bibinfo {year} {2013})}\BibitemShut {NoStop}%
	\bibitem [{\citenamefont {Serha}\ \emph {et~al.}(2022)\citenamefont {Serha}, \citenamefont {Bozhko}, \citenamefont {Agrawal}, \citenamefont {Verba}, \citenamefont {Kostylev}, \citenamefont {Vasyuchka}, \citenamefont {Hillebrands},\ and\ \citenamefont {Serga}}]{Serha2022}%
	\BibitemOpen
	\bibfield  {author} {\bibinfo {author} {\bibfnamefont {R.~O.}\ \bibnamefont {Serha}}, \bibinfo {author} {\bibfnamefont {D.~A.}\ \bibnamefont {Bozhko}}, \bibinfo {author} {\bibfnamefont {M.}~\bibnamefont {Agrawal}}, \bibinfo {author} {\bibfnamefont {R.~V.}\ \bibnamefont {Verba}}, \bibinfo {author} {\bibfnamefont {M.}~\bibnamefont {Kostylev}}, \bibinfo {author} {\bibfnamefont {V.~I.}\ \bibnamefont {Vasyuchka}}, \bibinfo {author} {\bibfnamefont {B.}~\bibnamefont {Hillebrands}},\ and\ \bibinfo {author} {\bibfnamefont {A.~A.}\ \bibnamefont {Serga}},\ }\bibfield  {title} {\bibinfo {title} {Low‐damping spin‐wave transmission in {YIG/Pt}‐interfaced structures},\ }\href {https://doi.org/10.1002/admi.202201323} {\bibfield  {journal} {\bibinfo  {journal} {Adv. Mater. Interfaces}\ }\textbf {\bibinfo {volume} {9}},\ \bibinfo {pages} {2201323} (\bibinfo {year} {2022})}\BibitemShut {NoStop}%
	\bibitem [{\citenamefont {B\"{u}ttner}\ \emph {et~al.}(2000)\citenamefont {B\"{u}ttner}, \citenamefont {Bauer}, \citenamefont {Demokritov}, \citenamefont {Hillebrands}, \citenamefont {Kivshar}, \citenamefont {Grimalsky}, \citenamefont {Rapoport},\ and\ \citenamefont {Slavin}}]{Buettner2000}%
	\BibitemOpen
	\bibfield  {author} {\bibinfo {author} {\bibfnamefont {O.}~\bibnamefont {B\"{u}ttner}}, \bibinfo {author} {\bibfnamefont {M.}~\bibnamefont {Bauer}}, \bibinfo {author} {\bibfnamefont {S.~O.}\ \bibnamefont {Demokritov}}, \bibinfo {author} {\bibfnamefont {B.}~\bibnamefont {Hillebrands}}, \bibinfo {author} {\bibfnamefont {Y.~S.}\ \bibnamefont {Kivshar}}, \bibinfo {author} {\bibfnamefont {V.}~\bibnamefont {Grimalsky}}, \bibinfo {author} {\bibfnamefont {Y.}~\bibnamefont {Rapoport}},\ and\ \bibinfo {author} {\bibfnamefont {A.~N.}\ \bibnamefont {Slavin}},\ }\bibfield  {title} {\bibinfo {title} {Linear and nonlinear diffraction of dipolar spin waves in yttrium iron garnet films observed by space- and time-resolved{B}rillouin light scattering},\ }\href {https://doi.org/10.1103/physrevb.61.11576} {\bibfield  {journal} {\bibinfo  {journal} {Phys. Rev. B}\ }\textbf {\bibinfo {volume} {61}},\ \bibinfo {pages} {11576–11587} (\bibinfo {year} {2000})}\BibitemShut {NoStop}%
	\bibitem [{\citenamefont {Dunagin}\ \emph {et~al.}(2025)\citenamefont {Dunagin}, \citenamefont {Serga},\ and\ \citenamefont {Bozhko}}]{Dunagin2025}%
	\BibitemOpen
	\bibfield  {author} {\bibinfo {author} {\bibfnamefont {R.~E.}\ \bibnamefont {Dunagin}}, \bibinfo {author} {\bibfnamefont {A.~A.}\ \bibnamefont {Serga}},\ and\ \bibinfo {author} {\bibfnamefont {D.~A.}\ \bibnamefont {Bozhko}},\ }\bibfield  {title} {\bibinfo {title} {Brillouin light scattering spectroscopy of magnon--phonon thermal spectra of an in-plane magnetized yig film in two-dimensional wavevector space},\ }\href {https://doi.org/10.1063/5.0251149} {\bibfield  {journal} {\bibinfo  {journal} {J. Appl. Phys.}\ }\textbf {\bibinfo {volume} {137}},\ \bibinfo {pages} {083901} (\bibinfo {year} {2025})}\BibitemShut {NoStop}%
	\bibitem [{\citenamefont {Sulymenko}\ \emph {et~al.}(2018)\citenamefont {Sulymenko}, \citenamefont {Prokopenko}, \citenamefont {Tyberkevych}, \citenamefont {Slavin},\ and\ \citenamefont {Serga}}]{Sulymenko2018}%
	\BibitemOpen
	\bibfield  {author} {\bibinfo {author} {\bibfnamefont {O.~R.}\ \bibnamefont {Sulymenko}}, \bibinfo {author} {\bibfnamefont {O.~V.}\ \bibnamefont {Prokopenko}}, \bibinfo {author} {\bibfnamefont {V.~S.}\ \bibnamefont {Tyberkevych}}, \bibinfo {author} {\bibfnamefont {A.~N.}\ \bibnamefont {Slavin}},\ and\ \bibinfo {author} {\bibfnamefont {A.~A.}\ \bibnamefont {Serga}},\ }\bibfield  {title} {\bibinfo {title} {Bullets and droplets: {T}wo-dimensional spin-wave solitons in modern magnonics ({R}eview {A}rticle)},\ }\href {https://doi.org/10.1063/1.5041426} {\bibfield  {journal} {\bibinfo  {journal} {Low Temp. Phys.}\ }\textbf {\bibinfo {volume} {44}},\ \bibinfo {pages} {602–617} (\bibinfo {year} {2018})}\BibitemShut {NoStop}%
	\bibitem [{\citenamefont {Donahue}\ and\ \citenamefont {Porter}(1999)}]{Donahue1999}%
	\BibitemOpen
	\bibfield  {author} {\bibinfo {author} {\bibfnamefont {M.~J.}\ \bibnamefont {Donahue}}\ and\ \bibinfo {author} {\bibfnamefont {D.~G.}\ \bibnamefont {Porter}},\ }\href@noop {} {\emph {\bibinfo {title} {OOMMF User's Guide, Version 1.0}}},\ \bibinfo {type} {Tech. Rep.}\ \bibinfo {number} {NISTIR 6376}\ (\bibinfo  {institution} {National Institute of Standards and Technology (NIST)},\ \bibinfo {address} {Gaithersburg, MD, USA},\ \bibinfo {year} {1999})\ \bibinfo {note} {available at \url{https://math.nist.gov/oommf/}}\BibitemShut {NoStop}%
	\bibitem [{\citenamefont {An}\ \emph {et~al.}(2013)\citenamefont {An}, \citenamefont {Vasyuchka}, \citenamefont {Uchida}, \citenamefont {Chumak}, \citenamefont {Yamaguchi}, \citenamefont {Harii}, \citenamefont {Ohe}, \citenamefont {Jungfleisch}, \citenamefont {Kajiwara}, \citenamefont {Adachi}, \citenamefont {Hillebrands}, \citenamefont {Maekawa},\ and\ \citenamefont {Saitoh}}]{An2013}%
	\BibitemOpen
	\bibfield  {author} {\bibinfo {author} {\bibfnamefont {T.}~\bibnamefont {An}}, \bibinfo {author} {\bibfnamefont {V.~I.}\ \bibnamefont {Vasyuchka}}, \bibinfo {author} {\bibfnamefont {K.}~\bibnamefont {Uchida}}, \bibinfo {author} {\bibfnamefont {A.~V.}\ \bibnamefont {Chumak}}, \bibinfo {author} {\bibfnamefont {K.}~\bibnamefont {Yamaguchi}}, \bibinfo {author} {\bibfnamefont {K.}~\bibnamefont {Harii}}, \bibinfo {author} {\bibfnamefont {J.}~\bibnamefont {Ohe}}, \bibinfo {author} {\bibfnamefont {M.~B.}\ \bibnamefont {Jungfleisch}}, \bibinfo {author} {\bibfnamefont {Y.}~\bibnamefont {Kajiwara}}, \bibinfo {author} {\bibfnamefont {H.}~\bibnamefont {Adachi}}, \bibinfo {author} {\bibfnamefont {B.}~\bibnamefont {Hillebrands}}, \bibinfo {author} {\bibfnamefont {S.}~\bibnamefont {Maekawa}},\ and\ \bibinfo {author} {\bibfnamefont {E.}~\bibnamefont {Saitoh}},\ }\bibfield  {title} {\bibinfo {title} {Unidirectional spin-wave heat conveyer},\ }\href {https://doi.org/10.1038/nmat3628} {\bibfield  {journal} {\bibinfo  {journal}
			{Nat. Mater.}\ }\textbf {\bibinfo {volume} {12}},\ \bibinfo {pages} {549–553} (\bibinfo {year} {2013})}\BibitemShut {NoStop}%
	\bibitem [{\citenamefont {Chumak}\ \emph {et~al.}(2009)\citenamefont {Chumak}, \citenamefont {Serga}, \citenamefont {Wolff}, \citenamefont {Hillebrands},\ and\ \citenamefont {Kostylev}}]{Chumak2009}%
	\BibitemOpen
	\bibfield  {author} {\bibinfo {author} {\bibfnamefont {A.~V.}\ \bibnamefont {Chumak}}, \bibinfo {author} {\bibfnamefont {A.~A.}\ \bibnamefont {Serga}}, \bibinfo {author} {\bibfnamefont {S.}~\bibnamefont {Wolff}}, \bibinfo {author} {\bibfnamefont {B.}~\bibnamefont {Hillebrands}},\ and\ \bibinfo {author} {\bibfnamefont {M.~P.}\ \bibnamefont {Kostylev}},\ }\bibfield  {title} {\bibinfo {title} {Scattering of surface and volume spin waves in a magnonic crystal},\ }\href {https://doi.org/10.1063/1.3127227} {\bibfield  {journal} {\bibinfo  {journal} {Appl. Phys. Lett.}\ }\textbf {\bibinfo {volume} {94}},\ \bibinfo {pages} {172511} (\bibinfo {year} {2009})}\BibitemShut {NoStop}%
	\bibitem [{\citenamefont {Wagle}\ \emph {et~al.}(2026)\citenamefont {Wagle}, \citenamefont {Stoeffler}, \citenamefont {Temdie}, \citenamefont {Kaffash}, \citenamefont {Castel}, \citenamefont {Majjad}, \citenamefont {Bernard}, \citenamefont {Henry}, \citenamefont {Bailleul}, \citenamefont {Jungfleisch},\ and\ \citenamefont {Vlaminck}}]{Wagle2026}%
	\BibitemOpen
	\bibfield  {author} {\bibinfo {author} {\bibfnamefont {D.}~\bibnamefont {Wagle}}, \bibinfo {author} {\bibfnamefont {D.}~\bibnamefont {Stoeffler}}, \bibinfo {author} {\bibfnamefont {L.}~\bibnamefont {Temdie}}, \bibinfo {author} {\bibfnamefont {M.~T.}\ \bibnamefont {Kaffash}}, \bibinfo {author} {\bibfnamefont {V.}~\bibnamefont {Castel}}, \bibinfo {author} {\bibfnamefont {H.}~\bibnamefont {Majjad}}, \bibinfo {author} {\bibfnamefont {R.}~\bibnamefont {Bernard}}, \bibinfo {author} {\bibfnamefont {Y.}~\bibnamefont {Henry}}, \bibinfo {author} {\bibfnamefont {M.}~\bibnamefont {Bailleul}}, \bibinfo {author} {\bibfnamefont {M.~B.}\ \bibnamefont {Jungfleisch}},\ and\ \bibinfo {author} {\bibfnamefont {V.}~\bibnamefont {Vlaminck}},\ }\bibfield  {title} {\bibinfo {title} {Shaping nonreciprocal caustic spin-wave beams},\ }\href {https://doi.org/10.1103/qz9q-bp9j} {\bibfield  {journal} {\bibinfo  {journal} {Phys. Rev. B}\ }\textbf {\bibinfo {volume} {113}},\ \bibinfo {pages} {L060405} (\bibinfo {year} {2026})}\BibitemShut
	{NoStop}%
	\bibitem [{\citenamefont {Holanda}\ \emph {et~al.}(2018)\citenamefont {Holanda}, \citenamefont {Maior}, \citenamefont {Azevedo},\ and\ \citenamefont {Rezende}}]{Holanda2018}%
	\BibitemOpen
	\bibfield  {author} {\bibinfo {author} {\bibfnamefont {J.}~\bibnamefont {Holanda}}, \bibinfo {author} {\bibfnamefont {D.~S.}\ \bibnamefont {Maior}}, \bibinfo {author} {\bibfnamefont {A.}~\bibnamefont {Azevedo}},\ and\ \bibinfo {author} {\bibfnamefont {S.~M.}\ \bibnamefont {Rezende}},\ }\bibfield  {title} {\bibinfo {title} {Detecting the phonon spin in magnon–phonon conversion experiments},\ }\href {https://doi.org/10.1038/s41567-018-0079-y} {\bibfield  {journal} {\bibinfo  {journal} {Nat. Phys.}\ }\textbf {\bibinfo {volume} {14}},\ \bibinfo {pages} {500–506} (\bibinfo {year} {2018})}\BibitemShut {NoStop}%
	\bibitem [{\citenamefont {Costa}\ \emph {et~al.}(2021)\citenamefont {Costa}, \citenamefont {Figeys}, \citenamefont {Sun}, \citenamefont {Van~Hoovels}, \citenamefont {Tilmans}, \citenamefont {Ciubotaru},\ and\ \citenamefont {Adelmann}}]{Costa2021}%
	\BibitemOpen
	\bibfield  {author} {\bibinfo {author} {\bibfnamefont {J.~D.}\ \bibnamefont {Costa}}, \bibinfo {author} {\bibfnamefont {B.}~\bibnamefont {Figeys}}, \bibinfo {author} {\bibfnamefont {X.}~\bibnamefont {Sun}}, \bibinfo {author} {\bibfnamefont {N.}~\bibnamefont {Van~Hoovels}}, \bibinfo {author} {\bibfnamefont {H.~A.~C.}\ \bibnamefont {Tilmans}}, \bibinfo {author} {\bibfnamefont {F.}~\bibnamefont {Ciubotaru}},\ and\ \bibinfo {author} {\bibfnamefont {C.}~\bibnamefont {Adelmann}},\ }\bibfield  {title} {\bibinfo {title} {Compact tunable {YIG}-based {RF} resonators},\ }\href {https://doi.org/10.1063/5.0044993} {\bibfield  {journal} {\bibinfo  {journal} {Appl. Phys. Lett.}\ }\textbf {\bibinfo {volume} {118}},\ \bibinfo {pages} {162406} (\bibinfo {year} {2021})}\BibitemShut {NoStop}%
	\bibitem [{\citenamefont {Whitney}\ \emph {et~al.}(2025)\citenamefont {Whitney}, \citenamefont {Lewis}, \citenamefont {Dickovick}, \citenamefont {Kalappattil}, \citenamefont {Carr},\ and\ \citenamefont {Wu}}]{Whitney2025}%
	\BibitemOpen
	\bibfield  {author} {\bibinfo {author} {\bibfnamefont {A.~G.}\ \bibnamefont {Whitney}}, \bibinfo {author} {\bibfnamefont {J.~M.}\ \bibnamefont {Lewis}}, \bibinfo {author} {\bibfnamefont {J.}~\bibnamefont {Dickovick}}, \bibinfo {author} {\bibfnamefont {V.}~\bibnamefont {Kalappattil}}, \bibinfo {author} {\bibfnamefont {L.~D.}\ \bibnamefont {Carr}},\ and\ \bibinfo {author} {\bibfnamefont {M.}~\bibnamefont {Wu}},\ }\bibfield  {title} {\bibinfo {title} {Propagation of spin waves in doubly periodic magnonic crystals},\ }\href {https://doi.org/10.1103/physrevapplied.23.064012} {\bibfield  {journal} {\bibinfo  {journal} {Phys. Rev. Appl.}\ }\textbf {\bibinfo {volume} {23}},\ \bibinfo {pages} {064012} (\bibinfo {year} {2025})}\BibitemShut {NoStop}%
	\bibitem [{\citenamefont {Levchenko}\ \emph {et~al.}(2026)\citenamefont {Levchenko}, \citenamefont {Davídková}, \citenamefont {Mikkelsen},\ and\ \citenamefont {Chumak}}]{Levchenko2026}%
	\BibitemOpen
	\bibfield  {author} {\bibinfo {author} {\bibfnamefont {K.~O.}\ \bibnamefont {Levchenko}}, \bibinfo {author} {\bibfnamefont {K.}~\bibnamefont {Davídková}}, \bibinfo {author} {\bibfnamefont {J.}~\bibnamefont {Mikkelsen}},\ and\ \bibinfo {author} {\bibfnamefont {A.~V.}\ \bibnamefont {Chumak}},\ }\bibfield  {title} {\bibinfo {title} {Review on spin-wave {RF} applications},\ }\href {https://doi.org/10.1109/tmag.2026.3657608} {\bibfield  {journal} {\bibinfo  {journal} {IEEE Trans. Magn.}\ ,\ \bibinfo {pages} {1–45}} (\bibinfo {year} {2026})}\BibitemShut {NoStop}%
\end{thebibliography}

%

\end{document}